\documentclass[10pt,letterpaper]{article}
\usepackage{opex3}

\begin{document}

\title{Time-cost analysis of a quantum key distribution system clocked at 100 MHz}

\author{X. F. Mo$^1$, I. Lucio-Martinez$^1$, P. Chan$^2$, C. Healey$^1$, S. Hosier$^3$, W.~Tittel$^1$}

\address{1. Institute for Quantum Information Science, and Department of Physics and Astronomy, University of Calgary, Calgary, Alberta, T2N 1N4, Canada \\ 2. ATIPS Labs, Department of Electrical and Computer Engineering, University of Calgary, Calgary, Alberta, T2N 1N4, Canada \\3. Southern Alberta Institute of Technology, Calgary, Alberta, T2M 0L4, Canada}

\email{ilucio@qis.ucalgary.ca} 



\begin{abstract*}
We describe the realization of a quantum key distribution (QKD) system clocked at 100 MHz. The system includes classical post-processing implemented via software, and is operated over a 12 km standard telecommunication dark fiber in a real-world environment. A time-cost analysis of the sifted, error-corrected, and secret key rates relative to the raw key rate is presented, and the scalability of our implementation with respect to higher secret key rates is discussed.
\\
\\
\end{abstract*}


\section{Introduction}

Quantum Key Distribution (QKD) takes advantage of the  peculiar quantum properties of single photons to distribute secret keys \cite{BB84,QC}. When implemented correctly \cite{Makarov10}, QKD, in combination with the One-Time Pad, allows two distant parties to communicate in an information-theoretic secure way over an untrusted but authenticated channel. 
 
A QKD system requires a quantum and a classical channel to distribute quantum information, here in form of quantum bits (qubits), and classical information, respectively. To obtain a secret key, a QKD system must complete the following steps:  1) Generation, faithful transmission, and measurement of qubits, yielding the \emph{raw key}. 2) Sifting of the raw key, i.e. comparison of the bases used by the sender and receiver to generate and detect each individual qubit. This is done over the classical channel. Only detection events where the bases match are kept, resulting in the \emph{sifted key}. 3) Error correction. The purpose of this step is to remove all errors in the sifted key due to a noisy channel or eavesdropping. This procedure requires communication over the classical channel. It yields information about the quantum bit error rate (QBER) of the sifted key and results in the {\it error-corrected key}. 4) Privacy amplification, the final step in QKD, shortens the error-corrected key and thereby removes all information that Eve might have obtained while eavesdropping. The result is the \emph{secret key}. Furthermore, the key exchange has to be authenticated to corroborate the identity of the authorized parties and to avoid a man-in-the-middle attack.  

For given loss in the quantum channel, the relevant figure of merit characterizing a QKD system is the secret key rate. Significant effort has been devoted over the past several years to increase this rate \cite{Dixon10}. However, with a few notable exceptions reporting actual rates up to 1 MHz \cite{Toshiba10, SECOQC}, the secret key rate is often calculated from the sifted key rate assuming a reasonable efficiency for error correction as compared to the Shannon limit \cite{Brassard93}, and taking into account a reduction of the error-corrected key during privacy amplification \cite{GLLP}. While this leads to a rate that has some predictive power, it states an upper bound that can only be attained if qubits are distributed continuously, key sifting, error correction and privacy amplification can keep up with the rate at which the raw key is obtained, and if the memory of the processor(s) in use can cope with the amount of data involved. These conditions may be difficult to satisfy in an actual system, in particular in the case of systems clocked at high rates.

In this paper we analyze the performance of our QKD system in view of a high secret key rate. The goal of the analysis is to determine the limitation on the key rate based on the time-cost of each of the steps mentioned above. We also propose improvements that we will pursue in the near future. The bottlenecks revealed in this analysis, while obtained using our QKD system, are likely to be relevant for other implementations as well. Hence, we believe that this study will help other research groups to develop high-rate QKD systems.

\section{Our QKD System}

\subsection{Hardware}
\begin{figure}
\centering \includegraphics[width=11cm]{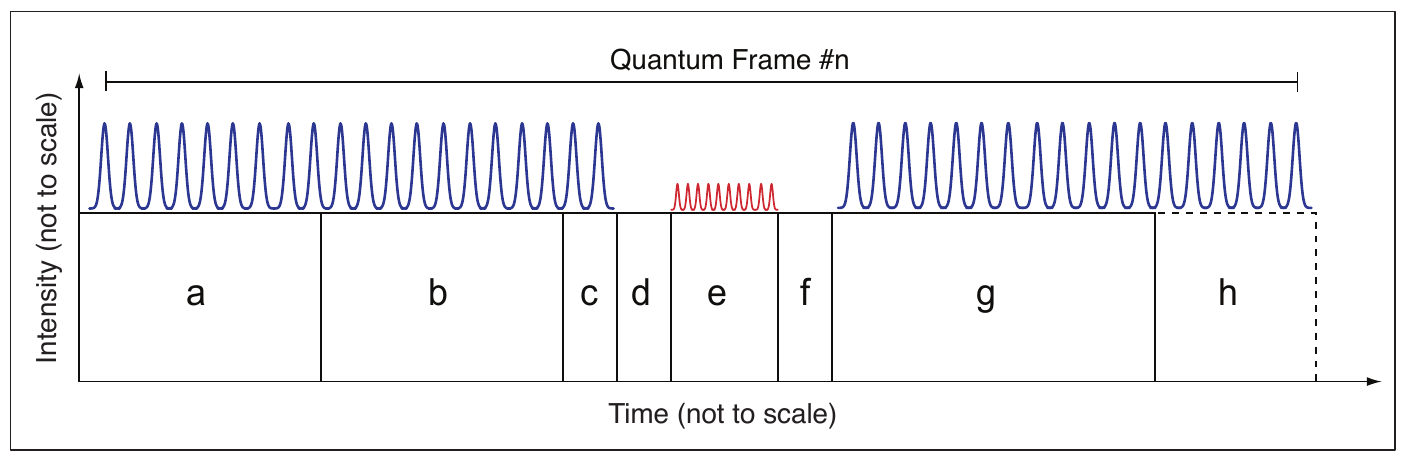}
\caption{Structure of the quantum frames. a: generation of bit and basis information to encode qubits; b: data transfer; c: generation of the classical control frame; d: deadtime, e: generation and transmission of quantum data (qubits); f:~deadtime, g: processing time, h: time for polarization control (when required).  \label{frames}}
\end{figure}

Our test took place between the  Quantum Cryptography and Communication Research Laboratory \mbox{(QCCRL)} at SAIT, where Alice is placed, and the Quantum Cryptography and Communication (QC2) Laboratory at the University of Calgary (UofC), where Bob is located. As usual, Alice and Bob denote the sender and receiver of quantum data, respectively. The transmission loss of the communication channel, a 12 km-long standard telecommunication fiber featuring many splices, is 6.5~dB. Our QKD system is fiber-based, implements the BB84 protocol supplemented with two decoy states \cite{Ma05} to detect photon number splitting attacks \cite{Dusek99, B00}, and employs polarization encoding. Furthermore, it is characterized by the use of quantum frames, which consist of alternating sequences of high-intensity laser pulses (forming classical control frames) and faint laser pulses (encoding quantum data), see figure~\ref{frames}. The classical control frames contain frame number and polarization information; the latter is used to assess and compensate time-varying birefringence in the communication channel \cite{lucio09}. The frames also contain information for clock synchronization and, in view of future integration into network environments, sender and receiver address to allow for routing. 

\begin{figure}
\centering \includegraphics[width=13cm]{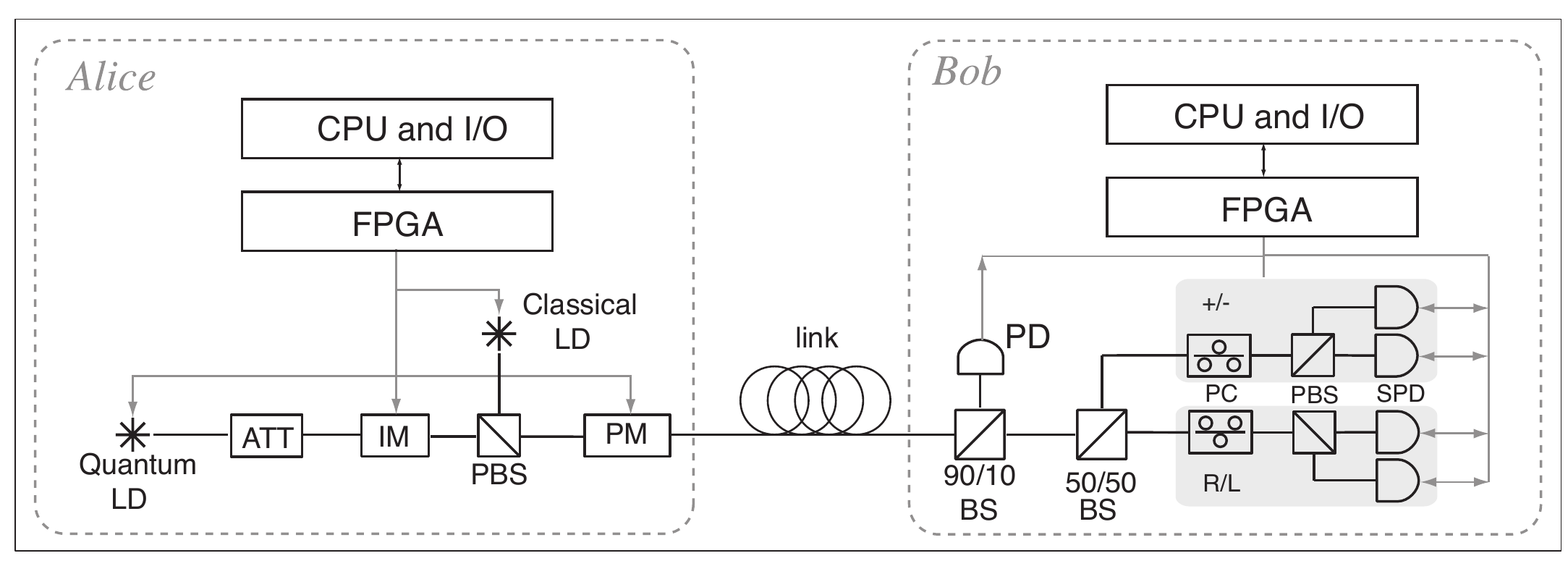}
\caption{Schematics of the optical and some electronic components of our QKD system. LD: laser diode; ATT: attenuator; IM: intensity modulator; PBS: polarization beam splitter;  PM: phase modulator; BS: beam splitter; PD: photo diode; PC: polarization controller; SPD: single photon detector; CPU: central processing unit (personal computer); FPGA: field programmable gate array; I/O: input/output interface. See text and \cite{lucio09} for more details.} \label{setup}
\end{figure}

Figure \ref{setup} shows a schematic of the optical and electronic components of our QKD system; a more detailed description of the optical part is given in \cite{lucio09}. Optical pulses of 500 ps duration and 1550 nm wavelength are generated by the \emph{quantum laser diode} and are attenuated using  a variable attenuator (ATT). To create the required signal and two decoy states, we use an intensity modulator (IM), generating weak pulses of light with mean photon numbers of $\mu$, 0.2$\mu$, 0.01$\mu$, respectively (the fixed relation between these three values is due to the way the attenuator and intensity modulator are used to generate loss). To encode the required polarization states, $\pm$ 45$^o$ linear polarized, and right- and left-circular polarized states, we use a polarization modulator (PM). Both modulators are configured to ensure passive compensation of temperature-dependent birefringence and polarization mode dispersion. On the receiver side, a photodiode is placed behind a 90/10 beamsplitter; it allows detecing the strong optical pulses, generated by the \textit{classical laser diode}, that form the control frames. Next, a 50/50 beamsplitter is placed to randomly select one of the two polarization bases for qubit measurement. Per basis, a voltage-controlled polarization controller (PC) and an optical detector (a low-bandwidth powermeter in the current system, not shown) are used  to compensate for time-varying polarization changes in the transmission line. This procedure relies on feedback from the classical control frames.

Polarization compensation executes whenever the QBER exceeds a certain threshold (between 3\% and 4.5\%, setup dependent). We have previously shown that the polarization stability over our real-world fiber link can vary greatly over time~\cite{lucio09}. Thus, for this feedback to work, the QBER must be updated sufficiently often, i.e. error correction must run on sifted key bits collected over a sufficiently short time\footnote{In the current setup, the number of sifted key bits to be processed in one execution of error correction is fixed to 10 kb. The time required to collect this data is setup dependent.}. In this case the feedback will ensure that the QBER is kept low when the channel is unstable (then generating only a small amount of raw key bits), while allowing key generation to run without interruption over several minutes during extended periods of stability.  The time needed for polarization compensation is determined by the reaction time of the powermeter, which limits the number of detectable voltage changes per second to one.

Qubit detection is either accomplished using four commercially available single photon detectors (SPDs) gated at 1 MHz, or using one high-rate, home-made detector \cite{Chris} that utilizes the self-differencing technique \cite{Yuan07} and allows photon detection up to 100 MHz. Currently, our QKD system is vulnerable to fake-state attacks \cite{Makarov10}, but preventive measures will be implemented in the near future. Note that qubit generation is clocked at 100 MHz in both cases.

\subsection{Software}
All  data is transferred via National Instruments digital I/O cards into or out of PCs with the following specifications; Alice:  AMD 64 X2 Dual Core 4600+, 2.4 GHz, 2 GB RAM, WinXP 32-bit; Bob:  Intel Core2 Quad CPU Q8300, 2.5 GHz, 4 GB RAM, Windows Vista 32-bit. Our system uses Field Programmable Gate Arrays (FPGAs) to control all active components. The clock rate, 100 MHz, is limited by the rate with which electronic signals are currently generated by the FPGA and can  be transmitted to, and converted by our home-made drivers that control the laser diodes and modulators. However, the optical components can generate qubits at a maximum rate of 980 MHz. Our system also includes classical post-processing (sifting, error correction and privacy amplification) implemented via software. Error correction is performed using low-density parity check codes (LDPC) \cite{Gallager, MacKay, Pearson}, and privacy amplification founds on Toeplitz matrices \cite{NTT}. 


Our QKD software is responsible for frame generation (Alice), data acquisition (Bob), key sifting, error correction, controlling polarization compensation, and writing collected data to the hard drives. The classical communication required for these tasks is performed using a TCP/IP connection established between the two computers over the public Internet. Each of the post-processing tasks can run independently, and both Alice and Bob run their tasks on one computer each.  The data gathered by the system is analyzed later on a computer with an Intel i5 CPU 760 @ 2.8GHz, where decoy state analysis and privacy amplification is performed.

The software is implemented primarily in National Instruments LabVIEW, with more time-intensive tasks being implemented in C++ libraries that are called as appropriate by the LabVIEW code.  These libraries may execute in parallel with any LabVIEW code that is not directly involved in controlling their execution. Hence, as more than one task may be executing at the same time, the elapsed times that we measure for processing tasks do not necessarily represent the required execution time. In particular, some tasks such as polarization compensation are not computationally intensive, but currently require significant time for the hardware to act.  During this time, computationally intensive tasks such as error correction may execute if there is data available to be processed.    

\section{System Performance}

To determine the performance of our QKD system, we increase the raw key rate from 0.1 to  120 kbps, and monitor the sifted and error-corrected key rates. As we will argue below, the secret key rate is simply related to the error-corrected rate by a factor described in \cite{GLLP}. As an example, the reduction assuming a QBER of 2.6\%, $\mu$=0.5 photons per pulse (with Poissonian photon number distribution), and no PNS attack is 23.5\% (see also \cite{lucio09}).

 In a first set of experiments, we employ four commercial single photon detectors gated at 1~ MHz. This effectively limits the clock rate of our QKD system to the same value. To change the raw key rate, we vary $\mu$ between 0.40 and 7.0 photons per pulse. Given the loss of 6.5~dB in the quantum channel, a detector efficiency of $\sim$10\%, as well as additional attenuation of  $\sim$3.5~dB in Bob's device, this yields raw key rates between $\sim$0.2  and 4.8 kbps.  While this procedure does not deliver secret keys for large values of $\mu$ (e.g. $\mu>1$), it does allow us to gauge how the system responds in the event of large raw key rates.  However, we point out that there is a limit to this procedure. Indeed, as $\mu$ increases, the probability that multiple detectors detect photons simultaneously also increases. This leads to larger processing requirements as only one, randomly selected detection is kept for subsequent steps \cite{Lutkenhaus00}. In turn, this leads to an underestimation of the sifted key, and hence error-corrected key rates (this effect was, however, not noticeable for $\mu\leq 7$). 

\begin{figure}
\centering \includegraphics[width=13cm]{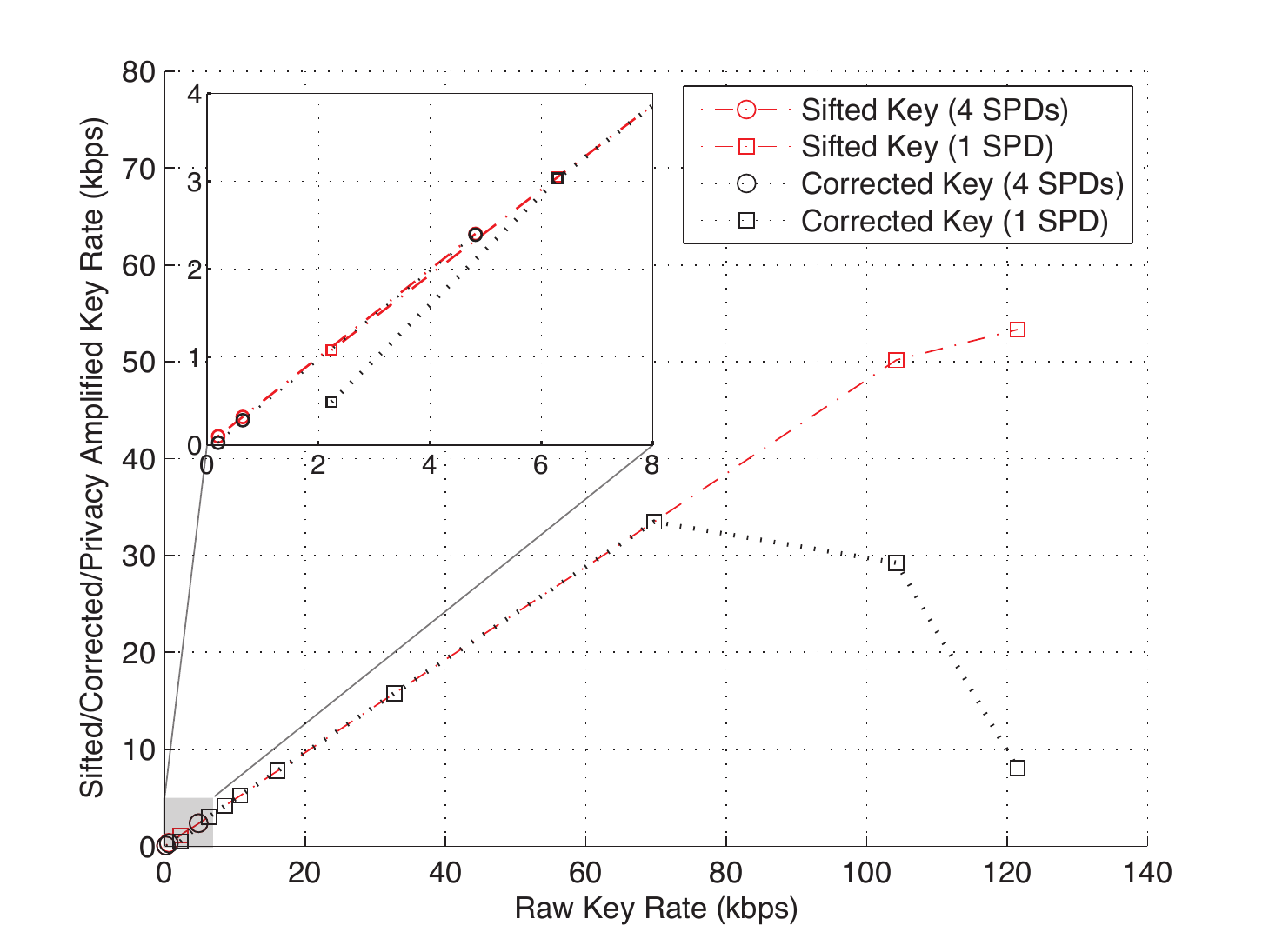}
\caption{Sifted and error corrected key rates as a function of the raw key rate. The inset shows the shaded area in more detail.}\label{ratepa}
\end{figure}

To obtain higher raw key rates, we perform a second set of measurements using a single, home-made SPD~\cite{Chris} that is gated at 100 MHz. We vary $\mu$ from 0.30 to 20 photons per pulse. Obviously, using only one detector does not allow distributing a secret key. Nevertheless, this setup allows increasing the raw key rate, and hence assessing the system performance in the event of large rates. More precisely, it delivers one quarter (i.e. 2.24 to 121 kbps) of the raw key rate we expect in a fully implemented QKD system with four high-rate detectors while providing a similar QBER.  All key rates listed below and in figure \ref{ratepa} refer to the actually detected (not extrapolated) rates.

%

Figure \ref{frames} shows the execution flow and frame structure in our system from Alice's perspective.  This perspective was chosen since Alice's timing currently limits  the maximum frame rate.  First, the state of all qubits within a quantum frame is determined by a software-based pseudo-random number generator (a, 225 ms)\footnote{Note that this solution is temporary - our final QKD system will employ physical random number generators.}.  This data is then transferred to a digital I/O card (b, 225 ms), which, along with an FPGA, controls our hardware. These devices generate the classical control frame (c, 960 ns), which includes a frame number, control information for polarization compensation, and a sender and receiver address that will be used for quantum packet routing in future work.  The header is followed by a deadtime (d, 50 ms), after which the qubits are generated and transmitted (e, 100 ms). A second deadtime (f, 50 ms) follows.  These deadtimes exist to avoid accidentally exposing the single photon detectors to strong light, which is generated at all times outside of the deadtimes and 'qubit time' (e). The second deadtime is followed by an idle time for the hardware, which is used by the computer for software post-processing and data logging (g, 55-130 ms, depending on the raw key rate). This time is determined by when the processor becomes available to generate the data for the next quantum frame. In particular, when the error rate has exceeded a certain threshold, the overall idle time is extended and then also comprises compensation for time-varying birefringence of the communication channel (h, averaging to 140 ms per frame). To summarize, qubits are transmitted on average during \mbox{100 ms} out of \mbox{845-920 ms}, i.e. during \mbox{10.9-11.8$\%$} of the system operation time.

The sifted and error corrected key rates obtained over the total system operation time are shown in figure \ref{ratepa} as a function of the raw key rate. From figure \ref{ratepa} we see that the error-corrected key rate peaks at 33.488 kbps at a raw key rate of 69.720 kbps, and that the sifted key rate does no longer increase linearly with respect to the raw key rate once the latter exceeds 114.160 kbps.  This is due to the fact that the post-processing software is run on the same computers as the data generation and collection software, and once processing resources are at their limit, the error correction, and subsequently key sifting, will get less execution time than is required to process all available data. To show the impact of this effect, error correction was run independently with simulated data, yielding a maximum error-corrected key rate of 53.213 kbps at an average QBER of 3.5$\%$. This average QBER is consistent with what is experienced during  operation of the QKD system with the four commercial single photon detectors and $\mu=1$. The QBER obtained using the high-rate detector is lower due to a better ratio between detection efficiency and dark count probability. In addition, the QBER decreases as $\mu$ is increased since the higher detection rates make dark counts less significant.  Thus, the 33.488 kbps obtained in the actual system is due to limited computational resources. Similarly, we also conclude that the sifted key rate is affected by process competition.

We have also performed decoy state analysis \cite{Ma05} on the data recorded by our QKD system to establish the maximum amount of information that may have leaked to an eavesdropper.  The execution time to process the data collected over 15 hours (enough to ignore finite key effects \cite{Rice09, Cai09}) was found to be less than a minute, and is thus negligible. Privacy amplification using the Toeplitz matrix approach has also been shown to require insignificant computational time if a number theoretic transform is used~\cite{NTT, Toshiba10}. Hence, the time required to establish and remove the eavesdropper's information does not need to be considered in our time-cost analysis.  In addition, while authenticated communication is typically used for sifting and error correction to prevent a man-in-the-middle attack \cite{Fung10}, it has been shown that this can be replaced by an authenticated verification of the secret key after privacy amplification~\cite{auth}. Hence, the impact of authentication on the secret key rate is negligible as well.


\section{Proposed Improvements}

The secret key rate of our QKD system can be improved by increasing the proportion of time spent transmitting quantum data.  Three simple modifications to our close to sequential execution of tasks stick out.  First, the deadtimes (d, f) can be shortened.  In principle, these times can be less than a millisecond.  However, as our system is still under development, we have chosen to maintain a large safety margin in case changes are made that alter the relative timing at the sender and receiver.  Second, we write more information to the hard drives than is necessary to perform decoy state analysis, privacy amplification, and authentication.  This increases the hardware idle time (g), but allows for a thorough analysis of the system. The secret key rate can thus be improved by writing only necessary information to file, or by replacing our current approach with a more efficient method of data transfer. Third, the powermeters used for polarization compensation have a response time on the order of a second, and multiple measurements are necessary to determine the necessary adjustments to the polarization controller.  As such, the polarization compensation time (h) can be reduced to a few ms by using fast photodetectors in conjunction with a fast polarization controller.  

In addition, we note that the 100~ms time interval for qubit transmission in each frame (e) is determined by the clock rate of 100~MHz and by the available memory in the digital I/O card, which allows generating $10^7$ qubits per frame. This limitation is present in our system for both detector setups, as Alice's system generates qubits at 100 MHz even when the detector gate rate is limited to 1~MHz.  While reducing Alice's clock frequency in the  case of the commercial detectors would bring the time used for qubit transmission much closer to 100$\%$ of the system operation time, this would ideally provide only a 8-9 fold increase in the raw key rate.  In comparison, using the fast detector provided more than a 60 fold increase in raw key rate.  In the case of the fast detector setup, it is possible to add more memory to the I/O card.  However, this would result in a proportional increase in the time required for the data preparation (a), data transfer (b), and key sifting plus error correction (g) steps.  This suggests that a faster interface to the computer, a faster random number generation,  as well as more efficient post-processing needs to be explored.

As used in QKD, LDPC encoding is not computationally intensive, requiring only a series of parity calculations~\cite{Pearson}.  Decoding, however, is an iterative process that uses the received data, parity information, and an initial estimate of the error rate (derived from previous executions of the protocol) in order to compute better estimates of the probability that each bit is in error~\cite{Gallager, MacKay}.  This iterative process ends successfully when the most likely result for the corrected data is consistent with the parity information, and failure is declared if a set maximum number of iterations is reached without meeting this condition.  In order to improve the throughput of the error correction in our system, two approaches are possible.  The computations required for LDPC decoding algorithm are well suited to parallel implementations.  Thus CPU utilization can be improved in our software implementation by taking advantage of this fact.  Moreover, LDPC decoding is well suited to hardware implementation~\cite{Levine}, and performing the decoding using specialized hardware, whether in an FPGA or custom integrated circuit, can yield error-corrected key rates of Mbps error-corrected key rates, as we have shown in~\cite{lucio09}. It should also be noted that the error correcting code used in our experiment was originally designed for use with the commercial detectors.  Since these detectors only provide a small key rate, a short block length of $10^4$ bits was used for the LDPC code in order to evaluate the QBER, and hence provide feedback to initiate the polarization control procedure in a timely fashion. The block length of the code can be increased significantly when using fast detectors, leading to better performance relative to the Shannon limit.  This, in turn, translates to a higher secret key rate since less information is revealed to the eavesdropper in this process.

Similarly, one should investigate hardware-based key sifting. In particular, executing sifting and error correction within the same FPGA would avoid data transfer into and out of a CPU, thereby avoiding time-consuming data transfer.

\section{Conclusions}

We have demonstrated a QKD system that implements the BB84 protocol supplemented with decoy states and quantum frames. The system executes software-based key sifting and error correction in real-time over a real-world fiber optic channel. We have done a time-cost analysis of all steps required in the generation of a secret key, and proposed improvements to our current implementation. Furthermore, we have analyzed the scalability of the sifted, error corrected and privacy amplified key rate with respect to the raw key rate, finding them to be determined by the sequential execution of the different steps in the key distribution protocol. Consequently, all processes that take significant time despite optimization have to be executed parallel to the distribution of qubits using dedicated, possibly custom hardware. Ignoring communication time, transmission loss and detector efficiency, the secret key rate would then be limited by the clock rate and the detector gate rate, i.e. 100 MHz in our current implementation with high-rate detectors.   

\section*{Acknowledgements}
The authors thank V. Kiselyov for technical support. This work is supported by General Dynamics Canada,
Alberta's Informatics Circle of Research Excellence (iCORE, now part of Alberta Innovates Technology Futures), the National Science and Engineering Research Council of Canada (NSERC), QuantumWorks, Canada Foundation for Innovation (CFI), Alberta Advanced Education and Technology (AET), and the Mexican Consejo Nacional de Ciencia y Tecnolog\'{\i}a (CONACYT).

\end{document}